\documentclass[hyper]{JHEP3} 

\usepackage{epsfig}

\newcommand\fverb{\setbox\pippobox=\hbox\bgroup\verb}

\newcommand\fverbdo{\egroup\medskip\noindent%

            \fbox{\unhbox\pippobox}\ }

\newcommand\fverbit{\egroup\item[\fbox{\unhbox\pippobox}]}

\newbox\pippobox

\title{Note About Integrability and
Gauge Fixing for Bosonic String
on $AdS_5\times S^5$}

\author{J. Kluso\v{n}
 \footnote{On leave from Masaryk University, Brno}\\
Dipartimento di Fisica \& Sezione I.N.F.N.\\
Universit\`a di Roma
``Tor Vergata'' \\
Via della Ricerca Scientifica 1 00133  Roma   ITALY\\
E-mail:
\email{Josef.Kluson@roma2.infn.it}}
\preprint{
}
\abstract{This short note is devoted to the study
of the  integrability of the   bosonic string on
$AdS_5\times S^5$ in the uniform light-cone gauge.
We  construct  Lax connection
for gauge fixed theory and we argue that it is flat.}


\def\tr{\mathrm{Tr}}
\def\tJ{\tilde{J}}

\newcommand{\hG}{\hat{G}}

\newcommand{\tx}{\tilde{x}}

\newcommand{\com}[1]{\left[#1\right]}

\newcommand{\talpha}{\tilde{\alpha}}

\newcommand{\tg}{\tilde{g}}

\newcommand{\hg}{\hat{g}}

\newcommand{\hJ}{\hat{J}}
\newcommand{\bM}{\mathbf{M}}

\newcommand{\tphi}{\tilde{\phi}}

\newcommand{\mY}{\mathcal{Y}}
\newcommand{\mZ}{\mathcal{Z}}

\newcommand{\bPhi}{\mathbf{\Phi}}
\begin{document}
\section{Introduction and Summary}
It is well known that the
  string sigma model on
$AdS_5\times S^5$ is classically
integrable\cite{Bena:2003wd}
\footnote{For some works considering
integrability of sigma model on $AdS_5\times
S^5$, see
\cite{Kluson:2007vw,Kluson:2007gp,Grassi:2006tj,
Dorey:2006mx,Gromov:2006dh,Alday:2005ww,Das:2005hp,Alday:2005jm,
Arutyunov:2005nk,Frolov:2005dj,Chen:2005uj,Alday:2005gi,Das:2004hy,
Arutyunov:2004yx,Beisert:2004ag,Kazakov:2004nh,Berkovits:2004jw,
Hatsuda:2004it,Kazakov:2004qf,Alday:2003zb,Vallilo:2003nx}}.
More precisely, the authors  
\cite{Bena:2003wd} found a
 Lax formulation of the equations
of motion for the classical Green-Schwarz
superstring that leads to the existence of 
an infinite tower of conserved
charges in the classical world-sheet theory. It
is important to stress that this Lax formulation
was derived for diffeomorphism invariant
and $\kappa$ symmetry invariant theory.

On the other hand it was shown recently in 
\cite{Dorey:2006mx} that this fact does not 
quite coincide with the standard 
definition of integrability. 
 Integrability 
in the standard sense requires not only the existence
of a tower of conserved charges 
but also requires that these charges
be in involution. In other words 
the conserved charges should Poisson 
commute with each other.  
The  analysis presented 
in \cite{Dorey:2006mx} explicitly 
demonstrated that for classical string
moving on $R\times S^3$ submanifold 
of $AdS_5\times S^5$ that the Poisson
brackets of conserved charges are in involution.
Further, in our recent paper
\cite{Kluson:2007vw}  we performed 
the  Hamiltonian  analysis of 
the same model on the world-sheet with
general metric. We
showed that in case when
 either the diffeomorphism invariance
of the world-sheet theory was preserved
or the components of the metric 
were fixed while the gauge symmetries
generated by  Virasoro generators
were not fixed the theory is integrable
in the sense advocated in  
\cite{Dorey:2006mx}.

The situation becomes  more involved
in case when the gauge fixing functions depend on 
the phase space variables. An example of
 such a gauge
is  \emph{uniform light-cone gauge}
\cite{Arutyunov:2006gs,Arutyunov:2005hd}
\footnote{For recent discussion of this
gauge, see for example
\cite{Klose:2006zd,Astolfi:2007uz}.}.
As the modest contribution to the  study
of the integrability of the gauge fixed
theory  we would like to present arguments that
further support the claim that the string theory
in uniform light-cone gauge is integrable.
We explicitly construct 
 Lax connection for bosonic sting on $AdS_5
 \times S^5$ in uniform light-cone gauge and
 we argue that this Lax connection is flat
\footnote{For some previous works
discussing the integrability of the
gauge fixed theory, see 
\cite{Das:2005hp,Alday:2005gi,Arutyunov:2004yx}.}.
These arguments are based on 
the T-duality  approach for the gauge fixing that
was introduced in 
\cite{Kruczenski:2004cn}. However before
we proceed to this approach we show 
how T-duality is implemented in   case of principal
chiral model and its relation to integrability
\footnote{For some  reviews of T-duality,
see \cite{Giveon:1994fu,Alvarez:1994dn,Giveon:1994mw}.},
following  
\cite{Gaberdiel:1995mx,
Alvarez:1993qi,Kiritsis:1993ju,Rocek:1991ps}.
We argue  that for some special examples
of principal chiral models T-dual models are
also integrable. Unfortunately we are not
able to answer the question of integrability
of T-dual of principal 
chiral model in the full general case.

Then we proceed to the case of the bosonic
string on $AdS_5\times S^5$, following
 formulation presented in 
\cite{Frolov:2005dj,Arutyunov:2004yx}. 
Our goal is to study the question
whether the theory  
\cite{Arutyunov:2004yx,Arutyunov:2006gs,
Arutyunov:2005hd,Astolfi:2007uz}
formulated in the uniform light-cone
gauge  is integrable
as well. We proceed in following way.
 In order to find the formulation of the
theory in the uniform light-cone gauge
we use the approach presented in
\cite{Klose:2006zd,Kruczenski:2004cn}
that is more convenient for the study
of the gauge fixed theory. 
On the other hand we  argue, following
\cite{Frolov:2005dj,Arutyunov:2004yx},
that due to the fact that the
 original Lax connection 
is not T-duality invariant we have to 
perform field redefinition 
that introduces new Lax connection
that is T-duality invariant. Using
this improved  
Lax connection we can define 
the Lax connection in T-dual background
when we use the map between original
and T-dual variables. As the next 
step we perform the gauge fixing
following \cite{Klose:2006zd,Kruczenski:2004cn}.
Then we  argue that the gauge
fixed theory possesses the Lax
connection that is flat.

The extension of this work is as follows.
It is straightforward to apply
an approach presented in this
paper to the case of the
full Green-Schwarz superstring, following 
very nice analysis presented in  
\cite{Alday:2005ww}. 
On the other hand the second extension
of this work is more involved. 
Even if we were able to find Lax connection
for gauge fixed theory 
the Poisson bracket of the 
spatial components of Lax
connection has not been determined yet.
While the calculation
of the Poisson bracket between
spatial components of Lax connection
is straightforward 
\cite{Kluson:2007vw,Dorey:2006mx} in
case of the gauge fixed action it
is much more difficult
\cite{Das:2005hp}. Moreover,  
the Poisson bracket derived
there does not seem to have 
the form presented in
\cite{Maillet:1985ek,Maillet:1985ec}.
While an existence of Lax connection
for gauge fixed theory implies
an existence of the infinite number of
conserved charges  the fact that
the Poisson bracket of Lax connection
 \cite{Das:2005hp} does not take the
 standard form implies that it is
 not clear that these charges are
 in involution. Clearly this issue
 deserves further study.

The organisation of this paper is
as follows. In the next section
(\ref{second}) we review
the derivation of T-duality rules
for sigma model. 
Then in section (\ref{third}) we
present similar calculation in 
case of principal chiral model defined
on group manifold. 
In section (\ref{fourth}) we turn
to the case of principal chiral model
that defines bosonic string on 
$AdS_5\times S^5$. We define
Lax connection that is invariant
under T-duality and then we find
Lax connection for the theory fixed
in uniform light-cone gauge.

\section{T-duality for Sigma Model}\label{second}
In this section we introduce standard
notation. We start
with the sigma model action that
describes the propagation of closed
string on the background with several
$U(1)$ isometries
\begin{eqnarray}\label{sigmaphi}
S&=&-\frac{\sqrt{\lambda}}{4\pi} \int d
\tau d\sigma \sqrt{-\gamma}[\gamma^{\alpha\beta}
\partial_\alpha \phi^i\partial_\beta
\phi^j g_{ij}-\epsilon^{\alpha\beta}
\partial_\alpha \phi^i\partial_\beta
\phi^j b_{ij}+\nonumber \\
&+& 2\partial_\alpha
\phi^i(\gamma^{\alpha\beta}u_{\beta,i}-
\epsilon^{\alpha\beta}v_{\beta,i})+
\mathcal{L}_{rest}] \ . \nonumber \\
\end{eqnarray}
As usual we have introduced the
effective string tension
$\frac{\sqrt{\lambda}}{2\pi}$ that is
identified with the 't Hooft coupling
in the AdS/CFT correspondence, $\gamma_{\alpha\beta}$
is worldsheet metric with Minkowski
signature  that in conformal gauge is
$\gamma^{\alpha\beta}=(-1,1)$ and
$\epsilon^{\alpha\beta}=\frac{\varepsilon^{\alpha\beta}}{
\sqrt{-\gamma}} \ ,
\varepsilon^{\tau\sigma}=
-\varepsilon^{\sigma\tau}=1$. Next
we assume that the action is invariant
under the $U(1)$ isometry
transformations that are geometrically
realised as shifts of the angle
variables $\phi^i \ , i=1,2,\dots,d$.
In other words the string background
contains the $d$-dimensional torus
$T^d$. The action (\ref{sigmaphi})
explicitly shows the dependence
on $\phi^i$ and their coupling to the
background fields $g_{ij}\ ,
b_{ij}$ and
$u_{\alpha,i},v_{\alpha,i}$. These
background fields are independent on
$\phi^i$ but can depend on other
bosonic  string
coordinates which are neutral under the
$U(1)$ isometry transformations.
Finally $\mathcal{L}_{rest}$ denotes
the part of the Lagrangian that depends
on other fields of the theory. 

As previous  discussion suggests the
 action (\ref{sigmaphi}) is
invariant under the constant shift of
$\phi^i$
\begin{equation}
\phi'^i(\tau,\sigma)=\phi^i(\tau,\sigma)+
\epsilon^i \ .
\end{equation}
 Corresponding  Noether currents have
the form
\begin{equation}\label{current}
J^\alpha_i=-\frac{\sqrt{\lambda}}{2\pi}
\sqrt{-\gamma}(\gamma^{\alpha\beta}
\partial_\beta \phi^j g_{ji}
-\epsilon^{\alpha\beta}\partial_\beta
\phi^jb_{ij}+\gamma^{\alpha\beta}u_{\beta,i}-
\epsilon^{\alpha\beta}v_{\beta,i}) \  
\end{equation}
and obeys the equation 
\begin{equation}\label{paju}
\partial_\alpha J^\alpha_i=0 \  
\end{equation}
as a consequence of the equations of 
motion. 

Now we are ready to study T-duality 
for this model. We closely  follow
 \cite{Giveon:1994mw}.
Let us start with the T-duality on a circle
parametrised by $\phi^1$. As the next
step we   gauge the shift  symmetry 
$\phi'^1=\phi^1+\epsilon^1$
so that $\epsilon^1$ is now
function of  $\tau,\sigma$.
If we require that the action is
invariant under the non-constant
transformation we have to 
introduce  the appropriate
gauge field $A_\alpha$ in such
a way that
\begin{equation}
\partial_\alpha \phi^1
\rightarrow (\partial_\alpha \phi^1
+A_\alpha)\equiv D_\alpha \phi^1 \ . 
\end{equation}
At the same time we add to the
action the term $\tphi^1 
\epsilon^{\alpha\beta}F_{\alpha\beta}
$ in order to assure that the
gauge field has trivial dynamics. 
The field $\tphi^1$ is corresponding
Lagrange multiplier.
Then we obtain the gauge invariant
action  
\begin{eqnarray}\label{sigmaphig}
S&=&-\frac{\sqrt{\lambda}}{4\pi} \int d
\tau d\sigma \sqrt{-\gamma}
[\gamma^{\alpha\beta}
D_\alpha \phi^1D_\beta
\phi^1 g_{11}
+2\gamma^{\alpha\beta}D_\alpha \phi^1\partial_\beta
\phi^a g_{1a}
+\gamma^{\alpha\beta}\partial_\alpha \phi^a
\partial_\beta
\phi^b g_{ab}
-\nonumber \\
&-&\epsilon^{\alpha\beta}
\partial_\alpha \phi^a\partial_\beta
\phi^b b_{ab}
-2\epsilon^{\alpha\beta}
D_\alpha \phi^1\partial_\beta
\phi^b b_{1b}
+\nonumber \\
&+&2D_\alpha
\phi^1(\gamma^{\alpha\beta}u_{\beta,1}-
\epsilon^{\alpha\beta}v_{\beta,1})+
2\partial_\alpha
\phi^a(\gamma^{\alpha\beta}u_{\beta,a}-
\epsilon^{\alpha\beta}v_{\beta,a})+
\tilde{\phi}^1\epsilon^{\alpha\beta}
F_{\alpha\beta}+
\mathcal{L}_{rest}] \ , \nonumber \\
\end{eqnarray}
where  $a,b=2,\dots, d$. 
Now thanks to the gauge invariance
we can fix the gauge $\phi^1=0$ so
that the action above takes the form
\begin{eqnarray}\label{sigmaphigf}
S&=&-\frac{\sqrt{\lambda}}{4\pi} \int d
\tau d\sigma \sqrt{-\gamma}
[\gamma^{\alpha\beta}
A_\alpha A_\beta
 g_{11}
+2\gamma^{\alpha\beta}A_\alpha
 \partial_\beta
\phi^a g_{1a}
+\gamma^{\alpha\beta}
\partial_\alpha \phi^a\partial_\beta
\phi^b g_{ab}
-\nonumber \\
&-&\epsilon^{\alpha\beta}
\partial_\alpha \phi^a\partial_\beta
\phi^b b_{ab}
-2\epsilon^{\alpha\beta}
A_\alpha \partial_\beta
\phi^b b_{1b}
+\nonumber \\
&+& 2A_\alpha
(\gamma^{\alpha\beta}u_{\beta,1}-
\epsilon^{\alpha\beta}v_{\beta,1})+
2\partial_\alpha
\phi^a(\gamma^{\alpha\beta}u_{\beta,a}-
\epsilon^{\alpha\beta}v_{\beta,a})+
\tilde{\phi}^1\epsilon^{\alpha\beta}
F_{\alpha\beta}+
\mathcal{L}_{rest}] \ . \nonumber \\
\end{eqnarray}
If we now integrate $\tilde{\phi}^1$
we obtain that $F_{\alpha\beta}=0$
and hence $A_\alpha=\partial_\alpha
\theta$. Inserting back to the action
(\ref{sigmaphigf}) we obtain
the original action (\ref{sigmaphi})
after identification $\theta=\phi^1$. 
On the other hand if we integrate out
$A_\alpha$  we obtain
\begin{equation}\label{Aphi}
A_\alpha=\frac{1}{g_{11}}
(-\partial_\alpha \phi^a g_{1a}
+\gamma_{\alpha\beta}\epsilon^{\beta\rho}
\partial_\rho\phi^a b_{1a}
-(u_{\alpha,1}-\gamma_{\alpha\beta}
\epsilon^{\beta\rho}
v_{\rho,1})-\gamma_{\alpha\beta}
\epsilon^{\beta\rho}
\partial_\rho\tilde{\phi}^1 ) \ . 
\end{equation}
Since we have argued that $A_\alpha$
can be related to the original
coordinate $\phi^1$ as $A_\alpha=
\partial_\alpha\phi^1$
the relation (\ref{Aphi}) implies following
relation between 
original and T-dual variables 
 $\phi^i$ and $\tphi^i$
\begin{eqnarray}\label{relphitphih}
\epsilon^{\alpha\rho}\partial_\rho
\tphi^1 &=&-\gamma^{\alpha\rho}
g_{11}\partial_\rho\phi^1
-\gamma^{\alpha\rho}
\partial_\rho \phi^a g_{1a}
+\epsilon^{\alpha\rho}
\partial_\rho\phi^a b_{1a}
-\gamma^{\alpha\rho}
u_{\rho,1}+\epsilon^{\alpha\rho} v_{\rho,1} \ , \nonumber
\\
\tphi^a&=&\phi^a \ . 
\end{eqnarray}
Now plugging the result
(\ref{Aphi})
 into the action
above we obtain the action 
equivalent to (\ref{sigmaphi})
\begin{eqnarray}\label{sigmaphid}
S&=&-\frac{\sqrt{\lambda}}{4\pi} \int d
\tau d\sigma \sqrt{-\gamma}
[\gamma^{\alpha\beta}
\partial_\alpha \tphi^i\partial_\beta
\tphi^j \tilde{g}_{ij}-\epsilon^{\alpha\beta}
\partial_\alpha \tphi^i\partial_\beta
\tphi^j \tilde{b}_{ij}+\nonumber \\
&+&2\partial_\alpha
\phi^i(\gamma^{\alpha\beta}\tilde{u}_{\beta,i}-
\epsilon^{\alpha\beta}\tilde{v}_{\beta,i})+
\tilde{\mathcal{L}}_{rest}] \ , \nonumber \\
\end{eqnarray}
where \cite{Buscher:1987sk,Buscher:1987qj}
\begin{eqnarray}\label{Brul}
\tilde{g}_{11}&=&\frac{1}{g_{11}} \ , \quad
\tilde{g}_{ab}=g_{ab}-
\frac{g_{a1}g_{1b}-b_{1a}
b_{1b}}{g_{11}} \ ,  \quad
\tilde{g}_{1a}=\frac{b_{1a}}{g_{11}} \ ,
\nonumber \\
\tilde{b}_{ab}&=&b_{ab}-\frac{g_{1a}b_{1b}
-b_{1a}g_{1b}}{g_{11}} \ , \quad
\tilde{b}_{1a}=\frac{g_{1a}}{g_{11}} \ ,
\quad  
\tilde{b}_{a1}=-\frac{g_{1a}}{g_{11}} \ , 
\nonumber \\
\tilde{u}_{\alpha,1}&=&\frac{v_{\alpha,1}}
{g_{11}} \ , 
\quad 
\tilde{v}_{\alpha,1}=\frac{u_{\alpha,1}}{g_{11}} \ , 
\nonumber \\
\tilde{u}_{\alpha,a}&=&u_{\alpha,a}-
\frac{g_{1a}u_{\beta,1}-b_{1a}v_{\alpha,1}
}{g_{11}} \ , \nonumber \\
\tilde{v}_{\alpha,a}&=&
v_{\alpha,a}-\frac{g_{1a}v_{\alpha,1}-
b_{1a}u_{\alpha,1}}{g_{11}} \ , 
\nonumber \\
\tilde{\mathcal{L}}_{rest}&=&
\mathcal{L}_{rest}-
\gamma^{\alpha\beta}
\frac{u_{\alpha,1}u_{\beta,1}-
v_{\alpha,1}v_{\beta,1}}
{g_{11}}+
\epsilon^{\alpha\beta}
\frac{u_{\alpha,1}v_{\beta,1}-
v_{\alpha,1}u_{\beta,1}}
{g_{11}} \ . \nonumber \\
\end{eqnarray}
These relations will be useful when
we discuss the gauge fixed form 
of the bosonic string on $AdS_5\times S^5$
in  section (\ref{fourth}).
On the other hand in the next section we perform
the same T-duality analysis for
the special case of the sigma
model that can be written in the
form of principal chiral model.

\section{T-Duality for Principal
Chiral model and Integrability}
\label{third}
Let us consider
the special case of the sigma
model action (\ref{sigmaphi})
that is known as  principal chiral model
\begin{equation}\label{princ}
S=-\frac{\sqrt{\lambda}}{4\pi}
\int d\sigma d\tau 
\sqrt{-\gamma}\gamma^{\alpha\beta}
K_{AB} J^A_\alpha J^B_\beta \ , 
\end{equation}
where 
\begin{equation}
J=G^{-1}dG=J^AT_A \ , 
\end{equation}
and where $G$ is a group element from the
group $\mathcal{G}$ and where $T_A$ are 
generators of corresponding algebra
$\mathbf{g}$ that obey following
relations
\begin{equation}
\com{T_A,T_B}=f_{AB}^C T_C \ , \quad
\tr (T_AT_B)=K_{AB} \ ,
\end{equation}
where $K_{AB}$ is invertible matrix
and where $f_{AB}^C=-f_{BA}^C$ are
structure constants of the algebra $\mathbf{g}$. 
 The
indices $A,B$ label components of the basis
$T_A$. 
If we parametrise the group element
with the fields $x^M$ we can write the
current $J_\alpha^A$ as
\begin{equation}
J_\alpha^A=
E^A_M\partial_\alpha x^M  \ . 
\end{equation}
Finally we  introduced the metric 
\begin{equation}
g_{MN}=
E_M^A K_{AB}E^B_N \ 
\end{equation}
defined on some target manifold
labelled with coordinates $x^M$. In this
interpretation $E_M^A$ are vielbeins of the
target manifold
\cite{Evans:1999mj}.
Note also that $E^A$ can be written
as
\begin{equation}
E^A=\tr (G^{-1}dG T_B)K^{BA}
\end{equation}
and hence the line element 
$ds^2$ can be written as
\begin{equation}
ds^2=\tr(G^{-1}dG G^{-1}dG) \ . 
\end{equation}
It is well known that the principal
chiral model (\ref{princ}) is integrable
\cite{Evans:1999mj}. More precisely,
we can find Lax connection for the
action (\ref{princ}) that is flat.
Further, we can argue that this model
possesses infinite number of
integrals of motion that are in 
involution \cite{Kluson:2007vw}.

Following \cite{Gaberdiel:1995mx}
we  now consider the case 
when  algebra $\mathbf{g}$ contains
 Cartan sub algebra 
\begin{equation}
T_i \ , \quad 
\com{T_i,T_j}=0 \ , \quad  i=1,\dots,d \ , \quad 
\tr (T_iT_j)=K_{ij} \ ,
\end{equation}
where $d$ is the rank of the algebra. 
Let us also parametrise the group element as
\begin{equation}\label{gel}
G=e^{\sum_{i=1}^d \alpha^iT_i} 
h \ . 
\end{equation}
Using (\ref{gel})  
we obtain
\begin{eqnarray}
& &\gamma^{\alpha\beta}
\tr (J_\alpha J_\beta)=
\gamma^{\alpha\beta}
\tr((h^{-1}\partial_\alpha \alpha^i  T_i h+
h^{-1}\partial_\alpha h)(
h^{-1}\partial_\beta \alpha^j T_jh+
h^{-1}\partial_\beta h))=
\nonumber \\
&=&\gamma^{\alpha\beta}
\partial_\alpha \alpha^i
K_{ij}\partial_\beta \alpha^j+
2\gamma^{\alpha\beta}
\partial_\alpha \alpha^i H_{i\beta}+
\gamma^{\alpha\beta}
\tr (h^{-1}\partial_\alpha h
h^{-1}\partial_\beta h) \ ,
\nonumber \\
\end{eqnarray}
where 
\begin{equation}
 H_{i\alpha}\equiv
\tr (T_i\partial_\alpha h h^{-1}) \ . 
\end{equation}
Now we are ready to study T-duality 
for this form of principal chiral model.
In order to have contact with the discussion
performed in next section let us consider
slightly more general case. Explicitly,
let us take 
 first two  $\alpha$'s
and consider following combination
\begin{equation}\label{alphaa}
\alpha^\alpha=\Gamma^\alpha_y\gamma^y \ , \quad  \alpha=1,2 \ ,
\quad 
x,y=1,2 \ , 
\end{equation}
where $\Gamma^x_y$ are  constant parameters.
Using (\ref{alphaa}) the action
(\ref{princ}) can be written as
\begin{eqnarray}\label{princ1}
S&=&-\frac{\sqrt{\lambda}}{4\pi}
\int d\sigma d\tau
\sqrt{-\gamma}
[\gamma^{\alpha\beta}
\partial_\alpha \gamma^1 \partial_\beta \gamma^1
K'_{11}+2\gamma^{\alpha\beta}\partial_\alpha \gamma^1
\partial_\beta\gamma^2 K'_{12}+
\gamma^{\alpha\beta}\partial_\alpha \gamma^2
\partial_\beta\gamma^2 K'_{22}+
\nonumber \\
&+&\gamma^{\alpha\beta}
\partial_\alpha \alpha^a
\partial_\beta \alpha^b K_{ab}
+2\gamma^{\alpha\beta}
\partial_\alpha \gamma^1 H'_{1\beta}+2
\gamma^{\alpha\beta}
\partial_\alpha \gamma^2 H'_{2\beta}+
\nonumber \\
&+&2\gamma^{\alpha\beta}\partial_\alpha \alpha^a
H_{a\beta}+
\gamma^{\alpha\beta}\tr(h^{-1}\partial_\alpha h
h^{-1}\partial_\beta h)] \ , 
\nonumber \\
\end{eqnarray}
where we  also 
presumed that the metric
$K_{ij}$ is block diagonal so
that $K_{\alpha a}=0 \ , a,b=3,\dots,d$.
In (\ref{princ1}) we  also
introduced the notation
\begin{equation}
K'_{xy}=\Gamma_x^\alpha\Gamma_y^\beta 
K_{\alpha\beta} \ , \quad 
H'_{x\alpha}=\Gamma_x^\beta  H_{\beta\alpha} \ .  
\end{equation}
Let us now consider $T$-duality along
the direction labelled with $\gamma^1$.
As in the previous section 
we gauge the theory corresponding 
to the shift of 
$\gamma^1$ 
\begin{equation}
\partial_\alpha\gamma^1\rightarrow
D_\alpha\gamma^1=\partial_\alpha\gamma^1
+A_\alpha \gamma^1 \ . 
\end{equation}
Then if we fix the gauge with 
 $\gamma^1=0$ the action takes the form
\begin{eqnarray}\label{princ2}
S&=&-\frac{\sqrt{\lambda}}{4\pi}
\int d\sigma d\tau
\sqrt{-\gamma}
[\gamma^{\alpha\beta}
A_\alpha A_\beta 
K'_{11}+2\gamma^{\alpha\beta}A_\alpha 
\partial_\beta\gamma^2 K'_{12}+
\gamma^{\alpha\beta}\partial_\alpha \gamma^2
\partial_\beta\gamma^2 K'_{22}+
\nonumber \\
&+&\gamma^{\alpha\beta}
\partial_\alpha \alpha^a
\partial_\beta \alpha^b K_{ab}
+2\gamma^{\alpha\beta}
A_\alpha H'_{1\beta}+\nonumber \\
&+&2\gamma^{\alpha\beta}\gamma^2 H'_{2\beta}+
2\gamma^{\alpha\beta}\partial_\alpha \alpha^a
H_{a\beta}+
\gamma^{\alpha\beta}\tr (h^{-1}
\partial_\alpha h h^{-1}
\partial_\beta h)
+\epsilon^{\alpha\beta}\tphi F_{\alpha\beta}] \ . 
\nonumber \\
\end{eqnarray}
If we integrate $\tphi$ we obtain
$\epsilon^{\alpha\beta}F_{\alpha\beta}=0$
that can be solved with
\begin{equation}
A_\alpha=\partial_\alpha \gamma^1 \ 
\end{equation}
and we recover the original action.
On the other hand if we integrate $A_\alpha$ we obtain
\begin{equation}
A_\alpha 
=-\frac{1}{K'_{11}}
[\partial_\alpha \gamma^1 K_{21}'
+H'_{1\beta}+
\gamma_{\alpha\gamma}\epsilon^{\gamma\beta}
\partial_\beta\tphi] \ . 
\end{equation}
Since $A_\alpha=\partial_\alpha \gamma^1$
this equation determines the relation between
original and dual variables
\begin{equation}\label{pargamma}
\partial_\alpha \gamma^1
=-\frac{1}{K'_{11}}
[\partial_\alpha \gamma^2 K_{21}'
+H'_{1\beta}+
\gamma_{\alpha\gamma}\epsilon^{\gamma\beta}
\partial_\beta\tphi] \ . 
\end{equation}
Inserting (\ref{pargamma}) into
(\ref{princ2}) 
 we obtain  dual action
\begin{eqnarray}\label{princT}
S&=&-\frac{\sqrt{\lambda}}{4\pi}
\int d\tau d\sigma
\sqrt{-\gamma}
[\gamma^{\alpha\beta}
\partial_\alpha\gamma^2\partial_\beta
\gamma^2(K'_{22}-\frac{K'^2_{12}}{K'_{11}})
+\gamma^{\alpha\beta}
\partial_\alpha \tphi\partial_\beta\tphi
\frac{1}{K'_{11}}
+\nonumber \\
&+&2\gamma^{\alpha\beta}
\partial_\alpha\gamma^2  
(H'_{2\beta}-\frac{K'_{21}}{K'_{11}}H'_{1\beta})
+\gamma^{\alpha\beta}
\partial_\alpha \alpha^a
\partial_\beta \alpha^b K_{ab}+
\nonumber \\
&+&2\gamma^{\alpha\beta}\partial_\alpha \alpha^a
H_{a\beta}+
\gamma^{\alpha\beta}(
\tr(h^{-1}\partial_\alpha h
h^{-1}\partial_\beta h)
-\frac{1}{K'_{11}}
H'_{1\alpha}H'_{1\beta})-
\nonumber \\
&-&\partial_\alpha \gamma^2
\partial_\beta \tphi \epsilon^{\alpha\beta}
\frac{2K'_{12}}{K'_{11}}
- \epsilon^{\alpha\beta}
\partial_\beta\tphi\frac{2H'_{1\alpha}}{K'_{11}}
] \ . 
\nonumber \\
\end{eqnarray}
Let us now observe that we can write
 \begin{eqnarray}
H'_{x\alpha}&=&\Gamma_{x}^\beta 
\tr (h^{-1}T_\beta h h^{-1}
\partial_\alpha h)=
\Gamma_x^\beta
\tr (h^{-1}T_\beta h T_A)
K^{AB}\tr (T_B h^{-1}\partial_\alpha h)=
\nonumber \\
&=&\Gamma_x^\beta
E_{\beta }^A K_{AB}E^B_m\partial_\alpha x^m=
\Gamma_x^\beta 
g_{\beta m}\partial_\alpha x^m\equiv
g'_{xm}\partial_\alpha x^m \ , 
\nonumber \\
K_{xy}'&=&\Gamma_x^\alpha \Gamma_y^\beta
\tr (T_\alpha T_\beta)=
\Gamma_x^\alpha \Gamma_y^\beta g_{\alpha\beta}
\equiv g'_{xy} \ 
\nonumber \\
 \end{eqnarray}
and consequently
\begin{eqnarray}
& & \gamma^{\alpha\beta}
\partial_\alpha \gamma^2
\partial_\beta \gamma^2
\left(K'_{22}-\frac{K'^2_{12}}{K'_{11}}
\right)=
\gamma^{\alpha\beta}
\partial_\alpha \gamma^2
\partial_\beta \gamma^2
\left(g'_{22}-\frac{g'_{12}g'_{12}}
{g'_{11}}\right) \ ,  \nonumber \\
& &\gamma^{\alpha\beta}
\partial_\alpha \gamma^2
\left(H'_{2\beta}-\frac{K'_{21}}{K'_{11}}
H'_{1\beta}\right)=
\gamma^{\alpha\beta}
\left(g'_{2m}-
\frac{g'_{12}g'_{1m}}{g'_{11}}\right)\partial_\alpha
\gamma^2\partial_\beta x^m \ , \nonumber \\
& &\gamma^{\alpha\beta}\left(
\tr(h^{-1}\partial_\alpha h
h^{-1}\partial_\beta h)
-\frac{1}{K'_{11}}
H'_{1\alpha}H'_{1\beta}\right)
=\gamma^{\alpha\beta}
\left(g'_{mn}-\frac{g'_{1m}g'_{1n}}
{g'_{11}}\right)\partial_\alpha x^m
\partial_\beta x^n \ , \nonumber \\
& &\partial_\alpha \gamma^2
\partial_\beta \tphi \epsilon^{\alpha\beta}
\frac{2K'_{12}}{K'_{11}}=
\partial_\alpha \gamma^2\partial_\beta \tphi
\epsilon^{\alpha\beta}
\frac{g'_{12}}{g'_{11}}
-\partial_\alpha\tphi\partial_\beta
\gamma^2 \epsilon^{\alpha\beta}
\frac{g'_{12}}{g'_{11}} \ , 
\nonumber \\
& &\epsilon^{\alpha\beta}
\partial_\beta\tphi\frac{2H'_{1\alpha}}{K'_{11}}=
\epsilon^{\alpha\beta}
\partial_\alpha x^m \partial_\beta
\tphi \frac{g'_{1m}}{g'_{11}}-
\epsilon^{\alpha\beta}
\partial_\alpha \tphi\partial_\beta
x^m\frac{g'_{1m}}{g'_{11}} \ .
\nonumber \\
\end{eqnarray}
In other words  T-dual action
(\ref{princT}) has exactly the same form
as the action 
(\ref{sigmaphid}) with the metric
and two form components given by
Buscher's rules (\ref{Brul})
when we replace $g_{MN}$ 
with $g'_{MN}$. 

Let us now 
restrict to the case when the
 metric $g'_{MN}$ is diagonal
and consequently $H'_{x\alpha}=0$. 
Then, in the similar way as in
\cite{Gaberdiel:1995mx} we
 introduce following generators
of the sub algebra of Cartan algebra
\begin{equation}
\tilde{T}_2=T_2'-\frac{K'_{12}}{K'_{11}}T_1' \ , \quad 
\tilde{T}_1=\frac{1}{K'_{11}}T_1' \ , \quad 
\tilde{T}_a=T_a
\end{equation}
and consider following group element
\begin{equation}
\tilde{G}=e^{\talpha^i \tilde{T}_i}h \ 
\end{equation}
with corresponding current
\begin{equation}
\tilde{J}=\tilde{G}^{-1}d\tilde{G}=
h^{-1}d\talpha^i \tilde{T}_i h+
h^{-1}dh  \ . 
\end{equation}
Then   we can  write T-dual action
(\ref{princT}) in the form
\begin{eqnarray}\label{princT1}
S=-\frac{\sqrt{\lambda}}{4\pi}
\int d\tau d\sigma
\sqrt{-\gamma}[\gamma^{\alpha\beta}
\tr \tJ_\alpha \tJ_\beta
-\partial_\alpha \gamma^2
\partial_\beta \tphi \epsilon^{\alpha\beta}
\frac{2K'_{12}}{K'_{11}}
] \ . 
\nonumber \\
\end{eqnarray}
Note that the last term can be written
as
\begin{eqnarray}
& &\frac{\sqrt{\lambda}}{2\pi}
\int d\sigma d\tau\partial_\alpha \gamma^2
\partial_\beta \tphi \varepsilon^{\alpha\beta}
\frac{2K'_{12}}{K'_{11}}=
\frac{\sqrt{\lambda}}{2\pi}\int d\sigma d\tau
\partial_\alpha
[\gamma^2\varepsilon^{\alpha\beta}\partial_\beta
\tphi \frac{2K'_{12}}{K'_{11}}]
-\nonumber \\
&-&\frac{\sqrt{\lambda}}{2\pi}
\int d\sigma d\tau
\gamma^2
\partial_\alpha[\varepsilon^{\alpha\beta}
\partial_\beta \tphi]
\frac{2K'_{12}}{K'_{11}}
\nonumber \\
\end{eqnarray}
and hence does not affect the equations
of motion. The first term is total derivative
and can be discarded from the action and
the second one vanishes due to the
antisymmetry of $\varepsilon^{\alpha\beta}$. 

The fact that T-dual action 
(\ref{princT1}) has again form of
the principal chiral model 
\footnote{Up to terms
that do not affect equations of motion.}
implies
that T-dual theory is integrable as well. 
On the other hand the form of the
group element  (\ref{gel})
is rather special.
For example, the principal chiral model that
describes bosonic string on $AdS_5\times S^5$
does not have such a simple form.  
\section{Integrability of Gauge 
Fixed Bosonic String 
on $AdS_5\times S^5$}\label{fourth}
The  motivation for the study 
of the question whether 
the integrability
of the principal chiral model is preserved
under T-duality  was  to understand the integrability
of the gauge fixed action for string
on $AdS_5\times S^5$. The problem 
is that this principal model does not
have such a simple form as an example
given in the end of the previous section
and hence we have to proceed in different
way. 

Explicitly, let us consider  action for
bosonic string on $AdS_5\times S^5$
 in the form
\begin{equation}\label{AdsP}
S=-\frac{\sqrt{\lambda}}{4\pi}
\int d\sigma d\tau
\sqrt{-\gamma}\gamma^{\alpha\beta}
g_{MN}\partial_\alpha x^M\partial_
\beta x^N \ ,  
\end{equation}
where $g_{MN}$ are metric components
of $AdS_5\times S^5$ whose explicit
form is given below and where $x^M$
label coordinates of this space. 

In order to study the integrability
properties of the theory we use
the fact that
 we can  write
the sigma model action (\ref{AdsP}) as
\cite{Arutyunov:2004yx}
\begin{equation}\label{AdSPr}
S=-\frac{\sqrt{\lambda}}{4\pi}
\int d\sigma d\tau
\sqrt{-\gamma}
\gamma^{\alpha\beta}
\tr (J_\alpha J_\beta)  \ , 
\end{equation}
where 
\begin{equation}
J_\alpha=G^{-1}\partial_\alpha G \ ,
\quad 
G=\left(\begin{array}{cc}
g_a & 0 \\
0 & g_s \\ 
\end{array}\right) \ . 
\end{equation}
Here $g_a$ and $g_s$
are  following $4\times
4$ matrices
\begin{equation}
g_a=\left(\begin{array}{cccc}
0 & \mZ_3 & -\mZ_2 &\mZ_1^* \\
-\mZ_3& 0 & \mZ_1 & \mZ_2^* \\
\mZ_2 & -\mZ_1 & 0 &-\mZ_3^* \\
-\mZ_1^* & -\mZ_2^* & \mZ_3^*
&0 \\
\end{array}\right) \ , \quad 
g_s=\left(\begin{array}{cccc}
0 & \mY_1 & -\mY_2 &\mY_3^* \\
-\mY_1& 0 & \mY_3 & \mY_2^* \\
\mY_2 & -\mY_3 & 0 & \mY_1^* \\
-\mY_3^* & -\mY_2^* & -\mY_1^*
&0 \\
\end{array}\right) \ ,
\end{equation}
where $\mZ_k, k=1,2,3$ are the
complex embedding coordinates for
$AdS_5$ and $\mY_k \ , k=1,2,3$
are the complex embedding coordinates
for sphere.  The matrix
$g_a$ is an element of the group 
$SU(2,2)$ since it can be shown that
\begin{equation}
g_a^\dag E g_a=E \ , \quad
E=\mathrm{diag}(-1,-1,1,1)
\end{equation}
provided the following condition is
satisfied
\begin{equation}
\mZ_1^*\mZ_1+\mZ_2^*\mZ_2
-\mZ_3^*\mZ_3=-1 \ .
\end{equation}
In fact $g_a$ describes embedding
of an element of the coset space
$SO(4,2)/SO(5,1)$ into group 
$SU(2,2)$ that is locally isomorphic to
$SO(4,2)$. We use this isometry to
work with $4\times 4$ matrices
rather with $6\times 6$ ones. Note
that due to the explicit choice of
the coset representative above there
is not any gauge symmetry left.
Quite analogously $g_s$ 
is unitary 
\begin{equation}
g_s g_s^\dag=1
\end{equation}
on condition that $\mY_1^*\mY_1+
\mY_2^*\mY_2+\mY^*_3\mY_3=1$.
The matrix $g_s$ describes 
an embedding of an element 
of the coset $SO(6)/SO(5)$
into $SU(4)$ being isomorphic
to $SO(6)$. 

The variables $\mZ,\mY$ are related
to the variables used in 
(\ref{AdsP})  as follows. 
 The five sphere $S^5$ is
parameterised by five variables: 
coordinates $y^i \ , i=1,\dots,4$ and
the angle variable $\phi$. In terms of
six real embedding coordinates $Y^A \ , 
A=1,\dots,6$ obeying the condition
$Y_AY^A=1$ the parametrisation reads
\begin{eqnarray}
\mY_1&=&Y_1+iY_2=\frac{y_1+iy_2}{
1+\frac{y^2}{4}} \ , \quad 
\mY_2=Y_3+iY_4=\frac{y_3+iy_4}{
1+\frac{y^2}{4}} \ , 
\nonumber \\
\mY_3&=&Y_5+iY_6=\frac{1-\frac{y^2}{4}}{
1+\frac{y^2}{4}}\exp (i\phi)
 \ .
\nonumber \\
\end{eqnarray}
In the same way we describe the $AdS_5$ 
space when we introduce four coordinates $z_i$ and
$t$. The embedding coordinates $Z_A$ that
obey $Z_AZ_B\eta^{AB}=-1$ with the metric
$\eta^{AB}=(-1,1,1,1,1,-1)$ is now parametrised
as 
\begin{eqnarray} 
\mZ_1&=&Z_1+iZ_2=-\frac{z_1+iz_2}{
1-\frac{z^2}{4}} \ , \quad 
\mZ_2=Z_3+iZ_4=-\frac{z_3+iz_4}{
1-\frac{z^2}{4}} \ , 
\nonumber \\
\mZ_3&=&Z_0+iZ_5=\frac{1+\frac{z^2}{4}}{
1-\frac{z^2}{4}}\exp (it)
 \ .
\nonumber \\
\end{eqnarray}
Note that the line element for
$AdS_5\times S^5$ takes the form
\begin{eqnarray}
ds^2=-\frac{(1+\frac{z^2}{4})^2}{(1-\frac{z^2}{4})^2}dt^2
+\frac{1}{(1-\frac{z^2}{4})^2}dz_idz_i+
\left(\frac{1-\frac{y^2}{4}}{1+\frac{y^2}{4}}\right)^2d\phi^2+
\frac{1}{(1+\frac{y^2}{4})^2}dy_idy_i \ . 
\nonumber \\ 
\end{eqnarray}
Now using the fact that
the bosonic string on $AdS_5\times
S^5$ can be written as principal
chiral model immediately implies
an existence of the Lax connection 
\begin{equation}\label{Laxor}
L_\alpha=\frac{1}{1-\lambda^2}
(J_\alpha-\Lambda \gamma_{\alpha
\beta}\epsilon^{\beta\gamma}J_\gamma) \ 
\end{equation}
that obeys the flatness condition
\begin{equation}\label{Laxflator}
\partial_\alpha L_\beta-\partial_\beta
L_\alpha+[L_\alpha,L_\beta]=0 \ . 
\end{equation}
Note that $\gamma_{\alpha\beta}$
in (\ref{Laxor}) is general
world-sheet metric and $\Lambda$
is a spectral parameter.

Then it was shown in  
\cite{Kluson:2007vw} that the
Poisson brackets of spatial
components of Lax connection
implies an existence of
infinite number of conserved
charges that are in involution.

On the other hand it would
be interesting to study the 
gauge fixed form of the theory
and whether the integrability
is preserved in this case. In fact
it was shown in 
\cite{Alday:2005gi,Arutyunov:2004yx}
that for some form of the gauge
fixing the theory is integrable
as well. Now we would like to give
an alternative argument that
supports the integrability of
the gauge fixed theory in 
uniform  light-cone gauge. 

Our approach is based on the
definition of the gauge fixing
introduced in  
\cite{Klose:2006zd,Kruczenski:2004cn}.
Let us introduce following 
combinations 
\begin{eqnarray}\label{defx+}
x^+=
(1-a)t+a\phi \ , \quad 
x^-=\phi-t \ , \nonumber \\
t=x^+-a x^- \ , \quad 
\phi=x^++(1-a)x^-  \ , 
\nonumber \\
\end{eqnarray}
where $a$ is a free parameter from interval
$a\in [0,1)$. Using these
variables 
 the action (\ref{AdsP})
takes the form
\begin{eqnarray}
S&=&-\frac{\sqrt{\lambda}}{4\pi}
\int d\sigma d\tau
\sqrt{-\gamma}\gamma^{\alpha\beta}
(g_{++}\partial_\alpha x^+
\partial_\beta x^++
2g_{+-}\partial_\alpha x^+
\partial_\beta x^-+\nonumber \\
&+& g_{--}\partial_\alpha x^-\partial_\beta x^-
+g_{mn}\partial_\alpha x^m
\partial_\beta x^n) \ , 
\end{eqnarray}
where now
\begin{equation}
g_{++}=g_{tt}+g_{\phi\phi} \ , 
\quad g_{+-}=
-ag_{tt}+(1-a)g_{\phi\phi} \ ,
\quad 
g_{--}=g_{tt}a^2+(1-a)^2g_{\phi\phi} \ . 
\end{equation}
and $x^m=(y_i,z_i)$. As the next
step  we perform T-duality along $x^-$.
Using  (\ref{relphitphih})
 we  obtain the relation between 
original and T-dual variables in the form
\begin{equation}\label{relot}
\epsilon^{\alpha\beta}
\partial_\beta \tx^-=\gamma^{\alpha\beta}
\partial_\beta x^+g_{+-}+
\gamma^{\alpha\beta}\partial_\beta x^-
g_{--} \ , \quad  \tx^m=x^m \ , \quad \tx^+=x^+ \ . 
\end{equation}
Then (\ref{relot}) also implies
\begin{equation}
\partial_\alpha x^-=
-\frac{1}{g_{--}}(\partial_\alpha \tx^+g_{+-}
+\gamma_{\alpha\beta}\epsilon^{\beta\gamma}
\partial_\gamma \tx^-) \ .
\end{equation}
Note  that formula's  (\ref{Brul})
also imply following forms of  metric an
two-form field components in 
T-dual theory
\begin{eqnarray}
\tg_{--}=\frac{1}{g_{--}}
=\frac{1}{g_{tt}a^2+(1-a)^2g_{\phi\phi}} 
 \ , \quad
\tg_{+-}=0 \ ,  \nonumber \\
\tg_{++}=g_{++}
-\frac{g_{-+}^2}{g_{--}}=
\frac{g_{tt}g_{\phi\phi}}
{g_{tt}a^2+(1-a)^2 g_{\phi\phi}}  
 \ , 
\nonumber \\
\tg_{mn}=g_{mn} \ , \quad  
\tilde{b}_{-+}=\frac{g_{-+}}{g_{--}}=
-\tilde{b}_{+-}=
\frac{-ag_{tt}+(1-a)g_{\phi\phi} }
{g_{tt}a^2+(1-a)^2g_{\phi\phi}} \ . 
\nonumber \\
\end{eqnarray}
As the next step we integrate
out the world-sheet metric
 $\gamma_{\alpha\beta}$
and we obtain
\begin{equation}\label{metriconshell}
\gamma_{\alpha\beta}=
\partial_\alpha \tx^M\partial_\beta
\tx^N \tg_{MN} \ .   
\end{equation}
Inserting this result to the T-dual
action we obtain
\begin{equation}
S=-\frac{\sqrt{\lambda}}{2\pi}
\int d\sigma d\tau\left[
\sqrt{-\det \tg_{MN}
\partial_\alpha \tx^M\partial_\beta
\tx^N}+\frac{1}{2}\varepsilon^{\alpha\beta}
\tilde{b}_{MN}\partial_\alpha 
\tx^M\partial_\beta\tx^N\right] \ .
\nonumber \\
\end{equation}
Finally, the uniform gauge fixing is
achieved as \cite{Klose:2006zd}
\begin{equation}\label{Klosefix}
\tx^+=\frac{\tau}{1-a} \ , \quad
\tphi=\frac{J_+\sigma}{2\pi} \ . 
\end{equation}
However since this approach is based
on T-duality transformation of the
action we come to the puzzle since
the Lax connection explicitly depends
on the variables that parametrise
 isometry directions. To resolve
this problem we follow 
\cite{Frolov:2005dj,Arutyunov:2004yx}.

We start with the  
original form of the action 
(\ref{AdSPr}) with general world-sheet
metric. 
Then we use the fact that matrices
$g_s,g_a$ enjoy following property
\cite{Frolov:2005dj,Arutyunov:2004yx}
\begin{eqnarray}
g_s(y,\phi)&=&
M(\phi)\hg_s(y)M(\phi)
\ , \nonumber \\
g_a(z,t)&=&
N(t)\hg_a(z)
N(t) \ , 
\nonumber \\
\end{eqnarray}
where
\begin{equation}
M(\phi)=
\left(\begin{array}{cccc}
e^{-\frac{i}{2}\phi} & 0 & 0 & 0 \\
0 & e^{\frac{i}{2}\phi} & 0 & 0 \\
0 & 0 & e^{\frac{i}{2}\phi} & 0 \\
0 & 0 & 0 & e^{-\frac{i}{2}\phi} \\
\end{array}\right) \ , \quad 
N(t)=
\left(\begin{array}{cccc}
e^{\frac{i}{2}t} & 0 & 0 & 0 \\
0 & e^{\frac{i}{2}t} & 0 & 0 \\
0 & 0 & e^{-\frac{i}{2}t} & 0 \\
0 & 0 & 0 & e^{-\frac{i}{2}t} \\
\end{array}\right) \ , 
\end{equation}
and
\begin{equation}
\hg_a=\left(\begin{array}{cccc}
0 & \frac{1+\frac{z^2}{4}}
{1-\frac{z^2}{4}} & -\mZ_2 &\mZ_1^* \\
-\frac{1+\frac{z^2}{4}}
{1-\frac{z^2}{4}}
& 0 & \mZ_1 & \mZ_2^* \\
\mZ_2 & -\mZ_1 & 0 &-
\frac{1+\frac{z^2}{4}}
{1-\frac{z^2}{4}} \\
-\mZ_1^* & -\mZ_2^* & \frac{1+\frac{z^2}{4}}
{1-\frac{z^2}{4}}
&0 \\
\end{array}\right) \ , \quad 
\hg_s=\left(\begin{array}{cccc}
0 & \mY_1 & -\mY_2 &
\frac{1-\frac{y^2}{4}}
{1+\frac{y^2}{4}}  \\
-\mY_1& 0 & 
\frac{1-\frac{y^2}{4}}
{1+\frac{y^2}{4}} & \mY_2^* \\
\mY_2 & -
\frac{1-\frac{y^2}{4}}
{1+\frac{y^2}{4}}
 & 0 & \mY_1^* \\
-\frac{1-\frac{y^2}{4}}
{1+\frac{y^2}{4}} & -\mY_2^* & -\mY_1^*
&0 \\
\end{array}\right) \ .
\end{equation}
Note that in this case
the matrix $G$ can be written
as
\begin{equation}\label{Gmhg}
G=\bM\hG \bM \ , \quad 
\bM=\left(\begin{array}{cc}
N(t) & 0 \\
0 & M(\phi) \\ \end{array}
\right) 
 \ , \quad 
 \hG=\left(\begin{array}{cc}
 \hg_a & 0 \\
 0 & \hg_s \\ \end{array}\right) \ . 
 \end{equation}
Using this factorisation property
 we obtain 
\begin{equation}\label{JM}
J=G^{-1}dG=\bM^{-1}(\hG^{-1}d\hG+\frac{i}{2}
\hG^{-1}d\bPhi \hG+\frac{i}{2}d\mathbf{\Phi})
\bM\equiv \bM^{-1}\hJ \bM \ , 
\end{equation}
where 
\begin{equation}
\mathbf{\Phi}=\left(\begin{array}{cc}
\Phi & 0 \\
0 & \Psi \\ \end{array}\right) \ ,
\end{equation}
and where $\Phi=\mathrm{diag}(-\phi,\phi,\phi,-\phi)$
and $\Psi=\mathrm{diag}(t,t,-t,-t)$. 
Now using (\ref{JM}) we define 
Lax connection $\hat{L}$ from the original
one (\ref{Laxor}) as  
\begin{equation}\label{hatL}
L_\alpha=\bM^{-1}\hat{L}_\alpha \bM
\end{equation}
Then the flatness condition
(\ref{Laxflator}) implies 
\begin{eqnarray}
& &\partial_\alpha L_\beta-\partial_\beta L_\alpha+
[L_\alpha,L_\beta]=\nonumber \\
&=&\bM^{-1}\left
(\partial_\alpha(\hat{L}_\beta-\frac{i}{2}
\partial_\beta\bPhi)-\partial_\beta(\hat{L}_\alpha
-\frac{i}{2}\partial_\alpha\bPhi)+\right.
\nonumber \\
&+& \left. \left[(\hat{L}_\alpha-\frac{i}{2}\partial_\alpha\bPhi),
(\hat{L}_\beta-\frac{i}{2}\partial_\beta\bPhi)\right]\right)\bM=0
\nonumber \\
\end{eqnarray}
Hence we see that instead of the original Lax connection
we can find another one that is again flat
\begin{eqnarray}
\mathbf{L}_\alpha &=&\hat{L}_\alpha(\bPhi)
-\frac{i}{2}\partial_\alpha\bPhi =\nonumber \\
&=&\frac{1}{1-\Lambda^2}
(\hat{J}_\alpha(\bPhi)-\Lambda
\gamma_{\alpha\beta}\epsilon^{\beta\gamma}
\hat{J}_\gamma(\bPhi))
-\frac{i}{2}\partial_\alpha\bPhi \ , \nonumber \\
\end{eqnarray}
where we explicitly stressed the dependence
of $\hJ$ on $\bPhi$ as follows from
(\ref{JM}). 
The advantage of the Lax connection $\mathbf{L}$
is that it now depends on a derivative of $\bPhi$ only.
This result implies that the Lax connection
$\mathbf{L}$ is useful for the definition of the
Lax connection for T-dual theory. Further,
since the relations between original
and T-dual variables are valid on-shell 
  we obtain
that the Lax connection defined
using the T-dual variables is flat
as well.
More precisely, using
(\ref{defx+}) we replace $t$ and $\phi$
in $\Phi,\Psi$  with $x^+,x^-$ so that 
\begin{eqnarray}
\Phi&=&(x^++(1-a)x^-) \Omega \ , \quad 
\Omega=\mathrm{diag}(-1,1,1,-1)  \ , 
  \nonumber  \\  
\Psi&=&(x^+-ax^-)\Sigma  \ , \quad 
\Sigma=\mathrm{diag}(1,1,-1,-1) \ .
\nonumber \\
\end{eqnarray}
Then using the relations between original
and T-dual variables 
(\ref{relot}) we  find
\begin{eqnarray}\label{Phii}
\partial_\alpha\Phi & = & [\partial_\alpha\tx^+-(1-a)
\frac{1}{g_{--}}(\partial_\alpha \tx^+g_{+-}
+\gamma_{\alpha\beta}\epsilon^{\beta\gamma}
\partial_\gamma \tx^-)]\Omega \ , 
\nonumber \\
\partial_\alpha\Psi & = & 
[ \partial_\alpha \tx^++a
\frac{1}{g_{--}}(\partial_\alpha \tx^+g_{+-}
+\gamma_{\alpha\beta}\epsilon^{\beta\gamma}
\partial_\gamma \tx^-)]\Sigma  \ .
  \nonumber \\
\end{eqnarray}
Then we can define Lax connection
for T-dual theory in the form
\begin{eqnarray}\label{LaxT}
\mathbf{L}_\alpha=\frac{1}{1-\Lambda^2}
(\hat{J}_\alpha(\bPhi)-\Lambda
\gamma_{\alpha\beta}\epsilon^{\beta\gamma}
\hat{J}_\gamma(\bPhi))
-\frac{i}{2}\partial_\alpha\bPhi \ ,  \nonumber \\
\end{eqnarray}
where now $\bPhi$ depends on $\tilde{x}^M$
through the relations (\ref{Phii}). Since (\ref{Phii})
hold on-shell
the Lax connection defined in T-dual theory (\ref{LaxT})
 is flat  as well. Note that we still presume
 that  the world-sheet 
metric $\gamma_{\alpha\beta}$ given in
(\ref{LaxT})  is general. However as
the next step in the gauge fixing
procedure  we integrate out it
and we get 
(\ref{metriconshell}). Again,
since Lax connection is flat for any metric it
is flat for metric 
that is  on-shell
(\ref{metriconshell}). 
Finally, we perform the gauge
fixing when we  insert  (\ref{Klosefix}) into
(\ref{Phii}) and we obtain components
$\partial \bPhi$ for gauge fixed theory
\begin{eqnarray}\label{Phiif}
\partial_\tau\Phi & = & [\frac{1}{1-a}-(1-a)
\frac{J_+}{2\pi}
\frac{\gamma_{\tau\tau}}{g_{--}
\sqrt{-\gamma}}]\Omega \ , 
\nonumber \\
\partial_\sigma\Phi & = & \frac{(1-a)J_+}{2\pi}
\frac{\gamma_{\sigma\tau}}{\sqrt{-\gamma}
g_{--}}\Omega \ , 
\nonumber \\
\partial_\tau\Psi & = & 
[ \frac{1}{1-a}+\frac{a J_+}{2\pi}
\frac{\gamma_{\tau\tau}}{
g_{--}\sqrt{-\gamma}}]\Sigma  \ .
  \nonumber \\
\partial_\sigma\Psi & = & 
-\frac{a J_+}{2\pi}
\frac{\gamma_{\sigma\tau}}
{\sqrt{-\gamma}
g_{--}}\Sigma  \ ,
  \nonumber \\
\end{eqnarray}
where now
\begin{eqnarray}
\gamma_{\tau\tau}
&=&\frac{1}{(1-a^2)}
\frac{g_{tt}g_{\phi\phi}}{g_{tt}
a^2+(1-a)^2g_{\phi\phi}}+g_{mn}
\partial_\tau x^m\partial_\tau x^n \ , 
\nonumber \\
\gamma_{\tau\sigma}&=&
g_{mn}\partial_\tau x^m
\partial_\sigma x^n  \ , \nonumber \\
\gamma_{\sigma\sigma}&=&
\frac{J_+^2}{4\pi^2(
g_{tt}a^2-(1-a)^2g_{\phi\phi})}+g_{mn}
\partial_\sigma x^m\partial_\sigma x^n \ . 
\nonumber \\
\end{eqnarray}
 As it is clear from
arguments given above the
 Lax connection for theory in the
uniform light-cone gauge
 (\ref{Klosefix}) is flat. We
 mean the study of the integrability
of the gauge fixed theory
 in the Lagrange formalism
can be considered as an useful
alternative to the analysis presented
in \cite{Alday:2005gi,Arutyunov:2004yx}.


\section*{Acknowledgements}

This work  was supported in part by the Czech Ministry of
Education under Contract No. MSM
0021622409, by INFN, by the MIUR-COFIN
contract 2003-023852 and , by the EU
contracts MRTN-CT-2004-503369 and
MRTN-CT-2004-512194, by the INTAS
contract 03-516346 and by the NATO
grant PST.CLG.978785.

\end{document}